\documentclass[preprint2]{aastex}
\usepackage{graphicx}
\slugcomment{Accepted in ApJ}
\lefthead{H. Murakami et al.}
\righthead{{\it Chandra} Observations of Diffuse X-Rays from the Sagittarius
B2 Cloud}
\begin{document}

\title{{\it Chandra} Observations of Diffuse X-Rays from the Sagittarius
B2 Cloud}
\author{Hiroshi Murakami, Katsuji Koyama}
\affil{Department of Physics, Faculty of Science, Kyoto University,
Sakyo-ku, Kyoto 606-8502, Japan; hiro@cr.scphys.kyoto-u.ac.jp,
koyama@cr.scphys.kyoto-u.ac.jp}
\and
\author{Yoshitomo Maeda\altaffilmark{1}}
\affil{Department of Astronomy and Astrophysics, The Pennsylvania State
University, 
525 Davey Lab. University park, PA 16802; maeda@astro.psu.edu}
\altaffiltext{1}{Subaru Telescope, National Astronomical Observatory of
Japan, 650 North Aohoku Place, Hilo, HI 96720}

\begin{abstract}
We present the first {\it Chandra} results of the
giant molecular cloud Sagittarius~B2 (Sgr~B2),
located about 100~pc away from the Galactic center. 
Diffuse X-rays are clearly separated from
one-and-a-half dozen resolved point sources.
The X-ray spectrum exhibits pronounced iron 
K-shell transition lines at 6.40~keV
(K$\alpha$) and 7.06~keV (K$\beta$), deep iron K-edge at 7.11~keV
and large photo-electric absorption at low energy.
The absorption-corrected X-ray luminosity is 
$\sim1\times10^{35}$~erg~s$^{-1}$, 
two orders of magnitude larger than the integrated luminosity
of all the resolved point sources.
The diffuse X-rays come mainly from the
south-west half of the cloud with a concave-shape pointing to the 
Galactic center direction.
These results strongly support 
the {\it ASCA} model that Sgr~B2 is
irradiated by an X-ray source at the Galactic center side.
\end{abstract}


\section{Introduction}
The central region of our Galaxy (here, GC) is behind the large
extinction in the optical band, but visible with the radio and
infrared bands.
The GC has been revealed to exhibit highly complex features, which are
possibly attributable to either a putative massive black hole (MBH) of
$\sim3\times10^6$ M$_\odot$ at the position of the radio non-thermal
source Sgr~A$^*$ \citep{Ghez00} or star burst activity in the GC.

The interstellar gas to the GC also becomes transparent
at higher energy X-rays.
Diffuse X-ray emission in the GC was discovered with the {\it
Einstein} Observatory \citep{Watson81}. The {\it Ginga}
satellite then found K$\alpha$ lines of highly ionized irons
associated
with a high temperature plasma in the central $\sim$100~pc region, 
the explosion energy of which is as large as $10^{54}$ ergs 
\citep{Koyama89,Yamauchi90}.
The {\it ASCA} satellite, with
the imaging capability in the wide energy X-ray band (0.5--10 keV)
coupled to the reasonable energy resolution, has further found a hot
plasma inside the Sgr~A radio shell with an oval-shape of $\sim
2'\times3'$ across \citep{Koyama96}.
Early results with the {\it Chandra} X-ray Observatory resolved out
numerous X-ray structures from the central 20~pc ($8'$) region of the
GC, and found a weak point source of $L_{\rm X} \simeq
10^{33}$~ergs~s$^{-1}$ at the
position of Sgr~A$^*$ \citep{Baganoff01}. This is nearly one billion times
lower than the Eddington luminosity for a MBH of
$\sim3\times10^6$M$_\odot$.

This "X-ray quiet" Sgr~A$^*$ is in a sharp contrast to the
high X-ray activity of the surrounding diffuse hot plasmas.
A hint to connect these
X-ray phenomena, the quiet Sgr~A$^*$ and active environment, was
obtained with the {\it ASCA} observations on Sgr~B2, a giant molecular
cloud located $\sim$100~pc away from Sgr~A$^*$. Koyama et al. (1996)
discovered diffuse emission of a K$\alpha$ 6.4 keV line of neutral or
low ionized irons from Sgr~B2. The X-ray spectrum and morphology of
Sgr~B2 are well explained by an "X-ray Reflection Nebula (XRN)" model;
the X-rays are due to reflection, photo-electric absorption and
fluorescence from iron atoms \citep{Murakami00}. However, no
adequately bright X-ray source was found in the vicinity of Sgr~B2 to
fully account for the diffuse X-ray flux. One likely scenario, despite
its considerable distance from Sgr~B2, is a past activity of
Sgr~A$^*$; an X-ray outburst of $L_{\rm X} \sim 3\times10^{39}$ ergs
s$^{-1}$ possibly caused by a surge of accretion onto the MBH in the
near past, and is currently in a quiescent accretion
\citep{Koyama96,Murakami00}.

The limited spatial resolution of {\it ASCA}, however, could not
exclude possible contamination of many point sources such as young
stellar objects in Sgr~B2, which might deform the spectrum and
morphology of the diffuse emission. We hence performed a high
resolution imaging spectroscopy on the Sgr~B2 region with {\it
Chandra}. We successfully detected diffuse X-rays, resolving many point
sources in Sgr~B2. In this paper, we focus on the diffuse X-rays,
while the results of the point sources will be given in a separate
paper. Throughout this paper, the distance to Sgr~B2 is assumed to be
8.5 kpc, the same to Sgr~A$^*$, and is within the error of estimated
distance of 7.1 $\pm$ 1.5~kpc \citep{Reid88}.

\section{Observation and Data Reduction}

The {\it Chandra} deep exposure observation on Sgr~B2 
was carried out on 29--30 March 2000. The satellite and
instrument are described by Weisskopf et al.\ (1996) and Garmire et
al.\ (2001) respectively. Sgr~B2 lies near the center of the ACIS
imaging array (ACIS-I) of four front-side illuminated CCDs, each with
$1024 \times 1024$ array of $0.5\arcsec \times 0.5\arcsec$ pixels
covering $8\arcmin.4 \times 8\arcmin.4$ field of view. Data
acquisition from ACIS was made in Timed-Exposure Faint mode with chip
readouts of every 3.24~s.

The ACIS front-side illuminated CCDs have suffered from an increase of
the parallel Charge Transfer Inefficiency (CTI), which decreases the
detection efficiency, energy gain and resolution. Sgr~B2 is
located at the top of amplifier 2--3 on chip I0, a location to be
heavily affected by the CTI. We therefore correct these effects using the
software described by Townsley et al.\ (2000), then no significant
variation is found in the energy of the instrumental Ni-K$\alpha$
(7.5 keV) and Au-L$\alpha$ (9.7 keV) lines (fluorescence due to particle
bombardment on the satellite metals) at the chip location of Sgr~B2 within
$<0.5$\% (90\% confidence). The energy resolution, however still varies
from row to row: 160 and 250~eV (FWHM) at the bottom and top of the chip
I0, respectively. In the spectral analysis, we use a response matrix
based on a nearly contemporaneous observation of reference lines from an 
on-board calibration source
(OBSID$=$62097), which are analyzed with the same CTI corrector. The
effective area of the telescope mirrors and the detection efficiency of
ACIS are calculated with the {\it mkarf} program in the {\it Chandra}
Interactive Analysis of Observations Software (CIAO, Version 1.0). 
To remove background events, we selected the
{\it ASCA} grades 0, 2, 3, 4, \& 6.  We also removed events from flaring
pixels and artificial stripes caused probably by hot pixels in the frame-store 
region and by particles which hit on the CCD node boundaries.

The total count rate over the chip is increased by 30\% in the middle
of the observation due to a high background rate caused by the impact
of solar energetic particles on the CCD chips. However, we do not
exclude the data during the high background rate because it does not
damage our imaging and spectroscopic studies. The effective exposure
time is 100~ks.

\section{Results and Analysis}
\subsection{Point Sources}
 
We first search foreground X-ray sources using the soft (0.5-1.5 keV)
band image. In the ACIS-I field of $17\arcmin.4 \times 17\arcmin.4$,
we find four bright X-ray sources, which have {\it Tycho-2}
counterparts. The positions of the {\it Tycho-2} sources have accuracy
of 60~mas \citep{Hog00}, hence the {\it Chandra} frame is fine-tuned
using these four sources within an absolute error of 0.25 arcsec
(1~sigma dispersion).

To highlight X-ray sources near or beyond the Sgr~B2 distance, we use
the 3.0--8.0 keV band image. The X-ray image near at Sgr~B2 is
complicated with diffuse emission and many faint point sources. For
the point-source search, we execute the CIAO `wavdetect' software of a
wavelet method \citep{Freeman00a}.  The threshold significances of
the wavdetect are set at $10^{-6}$ and 0.001 for the source list and
the background estimation, respectively. The wavelet scales used are
1, $\sqrt{2}$, 2, $2\sqrt{2}$, 4, $4\sqrt{2}$, 8, $8\sqrt{2}$, and 16
pixels. We then resolve about one-and-a-half dozen point sources in the
3.0--8.0 keV band from the central $3\arcmin \times 3\farcm5$ region
of the Sgr~B2 cloud. Neither radio nor infrared counterpart is found
from these sources, except possible association of the two brightest
sources to a compact HII complex near the center of the cloud, Sgr~B2
Main \citep{Vicente00}.

X-ray photons are extracted from a circle of a radius of six times of
the half power diameter (HPD) of point spread function (PSF) for each
source. Since the photon numbers of individual sources are limited, we
collect all the events from the point sources and make the combined
point-source spectrum, where the background spectrum is taken from the
point-source-free region in the Sgr~B2 cloud shown in Figure~1. The
background-subtracted combined spectrum is well fitted with an
optically thin thermal plasma model with the best-fit parameters shown
in Table~1. The integrated flux of all the point sources is 1.5
$\times 10^{-5}$ ph s$^{-1}$ cm$^{-2}$ in the 2--10 keV band, which is
about 8\% of the total diffuse flux (see \S 3.2). The mean $N_{\rm
H}$ value is estimated to be 1.9$\times 10^{23}$ H cm$^{-2}$, larger
than that of the Galactic absorption to Sgr~B2 of $\sim
1\times10^{23}$ H cm$^{-2}$ (Sakano et al. 2000), but smaller than
that through the Sgr~B2 cloud of $\geq$ 8$\times10^{23}$ H cm$^{-2}$
in the central 2$'$ radius region (Lis \& Goldsmith 1989, see also section
3.2). Therefore, we assume that most of the point sources are located
near or in the cloud (distance is 8.5 kpc) and estimate the integrated
luminosity (absorption corrected) in the 2--10 keV band to be
3$\times 10^{33}$ erg s$^{-1}$. Further details of individual point
sources will be given in a separate paper.

\subsection{Diffuse X-rays}

Since {\it ASCA} already found the strong 6.4 keV line from Sgr~B2, we
make the 6.4 keV-line (6.15-6.55 keV) image to highlight the most
detailed structure. Using the procedure given in Baganoff et
al. (2001) and the algorithm of Ebeling, White \& Rangarajan (2000),
the image is flat-fielded, and is smoothed adaptively with circular
Gaussian kernels. Figure~1 is the 6.4~keV line image overlaid on the
radio intensity contours of the molecule $^{13}$CO
\citep{Sato00}. The global morphology of the 6.4-keV emission is a
concave-shape located at the south-west side (right hand side) of the
molecular cloud pointing to the Galactic center direction (right hand
side). The positions of the resolved point sources mentioned in the
previous section are indicated by solid circles with radii of six
times of the HPD of PSF. No significant excess of the point sources is
found in this narrow band (6.15-6.55 keV) image, which indicates that
the 6.4 keV line of normal point sources, unlike the diffuse emission,
is not prominent. The local excess near the center and upper-left of
the molecular cloud are possibly due to HII complexes, Sgr~B2 Main at
$(l, b) \simeq (0.67, -0.04)$ and North at $(l, b) \simeq (0.68, -0.03)$. 
These two local excess occupy only 5~\% flux in the total diffuse flux.

The X-ray spectrum of the diffuse emission is made using the data from
the solid square and subtracting background taken from the dotted circle
in Figure~1. We then fit the spectrum with a phenomenological model, a
power-low function of photon index 2.0, adding K$\alpha$ and K$\beta$
lines of neutral iron and absorption with the Ba\l uci\'{n}ska-Church \&
McCammon (1992)\markcite{BalMc92} cross-sections for the solar
abundances \citep{Feldman92}. Iron abundance, which is sensitive to the
edge depth, is treated separately as a free parameter. The center energy
and flux of iron K$\alpha$ line are free parameters, but those of
K$\beta$ are linked to the K$\alpha$ parameters by fixing the ratio to the
laboratory values. Point source contamination is taken into account by
adding the best-fit model of the combined 
point-source spectrum (see \S 3.1).
The best-fit parameters are given in Table~2. We find that the
iron K$\alpha$ line energy is 
$6.38^{+0.02}_{-0.01}$~keV (errors adopted here and
after are 90\% confidence level unless otherwise noted), consistent with
neutral irons of 
6.40~keV. In order to examine whether highly
ionized irons are present or not, we add 6.70 keV line, K$\alpha$ of He-like
iron, and constrain that the equivalent width is less than 0.07 keV, or
at most 3 \% of that of neutral irons. Column density of neutral iron
($N_{\rm Fe}$) determined from the K-edge structure is $\sim 3.4 \times
10^{19}$ cm$^{-2}$, while $N_{\rm H}$ determined from the low energy
cutoff is $\simeq 8.8\times 10^{23}$ cm$^{-2}$. Both are nearly ten
times larger than those of the Galactic interstellar absorption, hence
are attributable to the gas in Sgr~B2. 
The {\it ASCA} results, in spite of its poor spatial resolution, are 
roughly consistent with the {\it Chandra} best-fit parameters,
supporting that the contribution of the point sources is almost
negligible (Murakami et al. 2000).

\section{Origin of the Diffuse X-rays}

Koyama et al. (1996) and Murakami et al. (2000) have argued that the
diffuse X-rays of Sgr~B2 can be explained by an XRN model; the Sgr~B2
cloud is irradiated by strong X-rays from an external source, absorbs
the X-rays, then reflects (Thomson scattering) and/or re-emits
fluorescent Fe lines. The discovery of diffuse 6.4 keV line from
other giant molecular clouds near the Radio Arc (Koyama et al. 1996)
and at Sgr~C (Murakami et al. 2001) may indicate that the presence of
XRNs at GC is common feature.

The present {\it Chandra} observation clearly detect diffuse 
X-rays from Sgr~B2 by resolving underlying point sources.
The diffuse X-rays are emitting from the
south-west half (right side in Figure~1) of the cloud 
with a concave-shape morphology 
pointing to the Galactic center direction. 
The spectrum is characterized with the pronounced 
iron K$\alpha$ line at 
$6.38^{+0.02}_{-0.01}$~keV, heavy absorption 
at iron K-edge and at low energy,
which confirm the {\it ASCA} results \citep{Koyama96,Murakami00}.
Furthermore, K$\beta$ line at 
$7.04^{+0.02}_{-0.01}$~keV is also discovered for the first time.
These results give further confirmation of the XRN concept of Sgr~B2.  
We, however, examine more critically for the origin of 
the diffuse X-rays based on the new observational facts.

\subsection{Young Stellar Objects}

The giant molecular cloud Sgr~B2 is known to be one of the richest
regions of ongoing star formation in the Galaxy (e.g.,
de Vicente et al. 2000). In fact, many compact HII regions as well as maser
sources have been found from the cloud. These sources trace the star
forming site in the cloud, and are found from the ridge running north
to south. Since young stellar objects (YSO), regardless low mass or
high mass stars, are rather strong X-ray sources with luminosity
ranging from 10$^{32-33}$ (massive YSO) to 10$^{28-29}$ erg~s$^{-1}$
(low mass YSO) (e.g., Feigelson \& Montmerle 1999), numerous unresolved YSOs in
Sgr~B2, if any, may mimic the diffuse X-rays. The spatial distribution
of the diffuse X-rays, a concave-shape, however is largely different
from that of the north-south ridge of star formation. The X-ray
spectra of YSOs are generally consistent with optically thin thermal
plasma of temperature ranging from 1 keV to a few keV, with the 6.7 keV line of
He-like irons present in some cases. The spectrum of the Sgr~B2 diffuse is
completely different from that of typical YSOs, hence a YSO origin is
ruled out for the bulk of the diffuse X-rays. Partial contribution of
YSOs may still be conceivable in the two local enhancement at the
cloud center and north, Sgr~B2 Main and North (see Figure~1). We hence
examine the X-ray spectrum from these two regions and find no
essential difference from the total diffuse spectrum. We only see a
hint of a weak 6.7 keV line with the equivalent width of 0.10$\pm$0.07
keV. Consequently, we can safely conclude that no significant
contribution of YSOs is found in the spectrum and morphology of the
diffuse X-rays of Sgr~B2.

\subsection{XRN Due to Internal Point Sources}

It may be conceivable that the resolved point sources in Sgr~B2
irradiate the surrounding gas and produce the XRN-like X-rays. These
point sources, however, are uniformly distributed in the cloud (see
Figure~1), which disagrees with the concave-shape of the diffuse
X-rays. The integrated luminosity of all the resolved point sources
is 3$\times$ 10$^{33}$ erg~s$^{-1}$, which is only 3\% of the X-ray
luminosity of the XRN. Therefore internal point sources can be ignored
in the XRN scenario.

\subsection{XRN Simulation by an External X-ray Source}

The diffuse X-rays are located on the south-west half of Sgr~B2, the
Galactic center side with the concave-shape pointing to the Galactic
center. These naturally lead us to suspect that an irradiating source 
should be located in the Galactic center direction. More quantitatively, 
we numerically simulate the XRN model with the assumption that the
gas distribution of Sgr~B2 is an elliptical cylinder-shape with the same 
radial dependence given in the model by Murakami et al. (2000) and is 
irradiated by an external source at the Galactic center direction.

In the XRN scenario, X-rays above the 7.11~keV energy are absorbed
mainly by iron atoms, while the lower energy X-rays are absorbed
mainly by lighter elements such as carbon, nitrogen and oxygen.
Continuum X-ray flux by the Thomson scattering is proportional to the
number of electrons ($\sim$hydrogen atoms). Accordingly, the ratio of
absorption depth at 7.11~keV and low energy band is proportional to
the abundance ratio of iron and the other elements. The equivalent
width of the 6.40~keV line constrains the iron abundance. We thus
simulate the spectrum with various abundance sets of iron and the
lighter elements and compare with the observed spectrum. The best-fit
spectrum is obtained with the reduced $\chi^2$ of 1.14 with 21 degrees
of freedom (d.o.f.). The result is given in Figure~2 by solid
histogram together with the data (crosses), where the dashed line is
contribution of background point sources. The abundances of iron and
lighter elements are 2.1$^{+0.9}_{-0.6}$ and 3.5$^{+2.0}_{-0.7}$ solar, 
respectively, consistent with, but more accurate than the {\it ASCA} results
(Murakami et al. 2000).

Figure~3 is a simulated XRN image of the 6.4~keV line
with the best-fit abundances.  We see a clear concave-shape emission, 
similar to the observed result. 
For quantitative comparison, we clearly need  more 
complicated assumption of gas distribution, which is beyond 
the scope of this paper.

Using the simulation, we estimate the luminosity of an
irradiating source should be $\sim 10^{39}$ ($D$/100pc)$^{2}$ erg
s$^{-1}$, in order to fully account for the flux of the diffuse X-rays,
where $D$ is a distance from the cloud center to the irradiating source.
No adequately bright X-ray source, however has been found near the
Sgr~B2 cloud (this work) nor at the central 20~pc region from Sgr~A$^*$
(e.g., Baganoff et~al. 2001). This strengthens the idea proposed by Koyama
et al. (1996) and Murakami et al. (2000) that a past activity of the
Galactic center is responsible for the presence of reflected X-rays and
absence of an irradiating source. The past activity of Sgr~A$^*$ would
be generated by an accretion surge on the MBH due, for example, to the
passage of dense shell of a young supernova remnant Sgr~A East as is
discussed by Maeda et al. (2001) based on the new {\it Chandra}
results of the GC.

\section{Summary}

1. Diffuse X-rays are clearly detected from the giant  molecular cloud Sgr~B2,
by resolving one and half dozens of  point sources. 

2. The morphology of the diffuse X-rays is concave-shape at 
the Galactic center side and  pointing to this direction.

3. The  spectrum exhibits strong emission lines at 6.40 and 7.06
keV, a large absorption at low energies below 4 keV, equivalent to
$9\times 10^{23}$ H cm$^{-2}$ (with the solar abundances) and a
pronounced edge-structure at 7.11 keV with equivalent $N_{\rm Fe}$ of 3
$\times 10^{19}$ cm$^{-2}$.

4. The resolved point sources  are likely located near or in  the cloud. 
The integrated luminosity of all the point sources is 3 $\times$ 10$^{33}$
erg~s$^{-1}$, which is only 3~\% of the diffuse X-rays.

5. The spectrum and morphology of the diffuse X-rays can not be
attributable to young stellar objects in the cloud, nor the
fluorescent by the resolved point sources.

6. The {\it Chandra} results strengthen the X-ray Reflection Nebula (XRN)
model, in which a strong X-ray source at the Galactic center side 
irradiates the Sgr~B2 cloud.

7. No luminous X-ray source to produce the XRN is found near the 
Sgr~B2 cloud. Possible idea, as already proposed by Koyama et al. (1996)
and Murakami et al. (2000), is that the Galactic center  Sgr~A$^*$  was active 
in the recent past but is quiescent at present.

\vspace*{1em}

\acknowledgements

The authors express their thanks to Eric Feigelson and Gordon Garmire
for kind hospitality to the {\it Chandra} data analysis at PSU. Our
study is technically supported by Margarita Karovska, a CXC support
scientist.  H. M. and Y. M. are financially supported by the Japan
Society for the Promotion of Science.

\clearpage

\bibliography{apj-jour}

\newpage
\onecolumn
\begin{figure}
\epsscale{0.5}
\includegraphics[angle=270,scale=0.7]{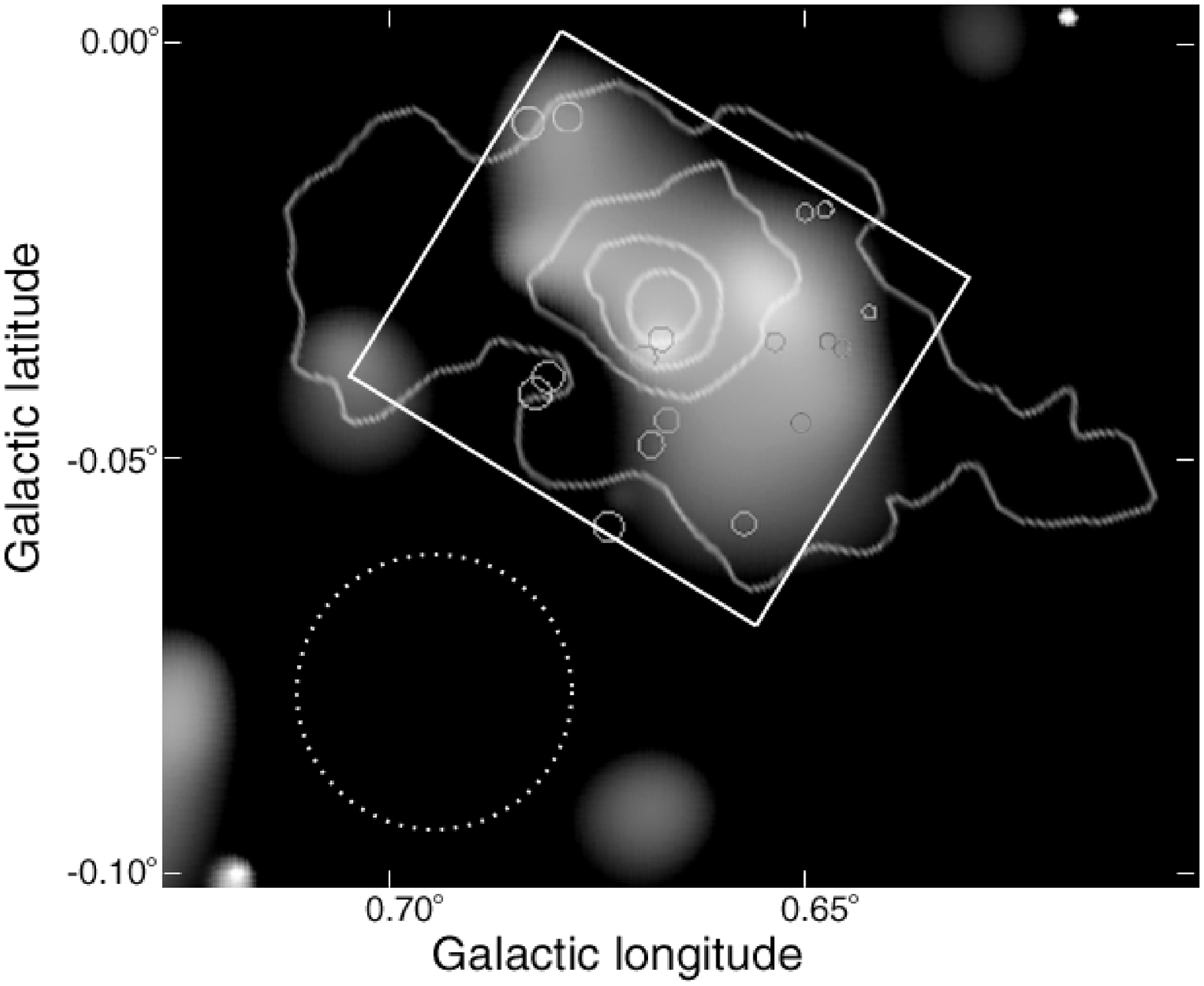} 
\caption[f1.ps]
{Adaptively smoothed ACIS-I image in the
6.15--6.55~keV band. The Sgr~B2 molecular cloud is indicated by the
contours of the $^{13}$CO line flux (Sato et al. 2000). The diffuse
X-rays are seen in the south-west side of Sgr~B2 (right side in the
figure) with a concave-shape pointing to the Galactic center
direction. The spectra of the diffuse X-rays and background are taken
from the solid rectangle ($3\arcmin \times 3\farcm5$) and the dotted
circle, respectively. The positions of the resolved point sources and
the data area for the combined spectrum are shown by the small solid
circles with the radius of six times of the HPD of PSF.}
\end{figure}

\begin{figure}
\epsscale{0.7}
\includegraphics[angle=270,scale=0.7]{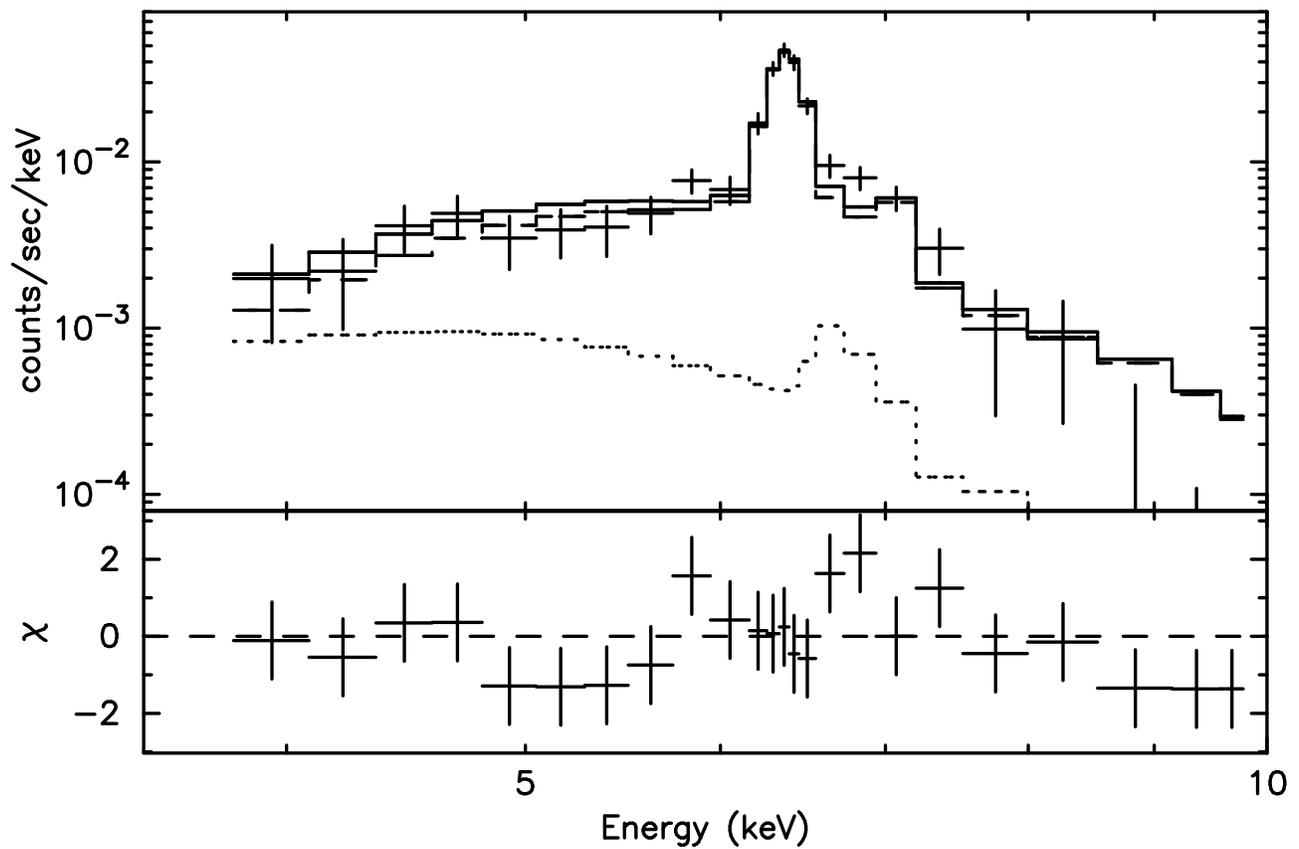} 
\caption[f2.ps]
{The spectrum of the diffuse X-rays from the Sgr~B2
cloud and the best-fit XRN model convolved with the response function (solid 
line). Residuals are shown in the bottom panel. 
The dashed line indicates the contribution of all the resolved point 
sources in Sgr~B2 (see Figure~1).}
\end{figure}

\begin{figure}
\epsscale{0.8}
\includegraphics[angle=270,scale=0.7]{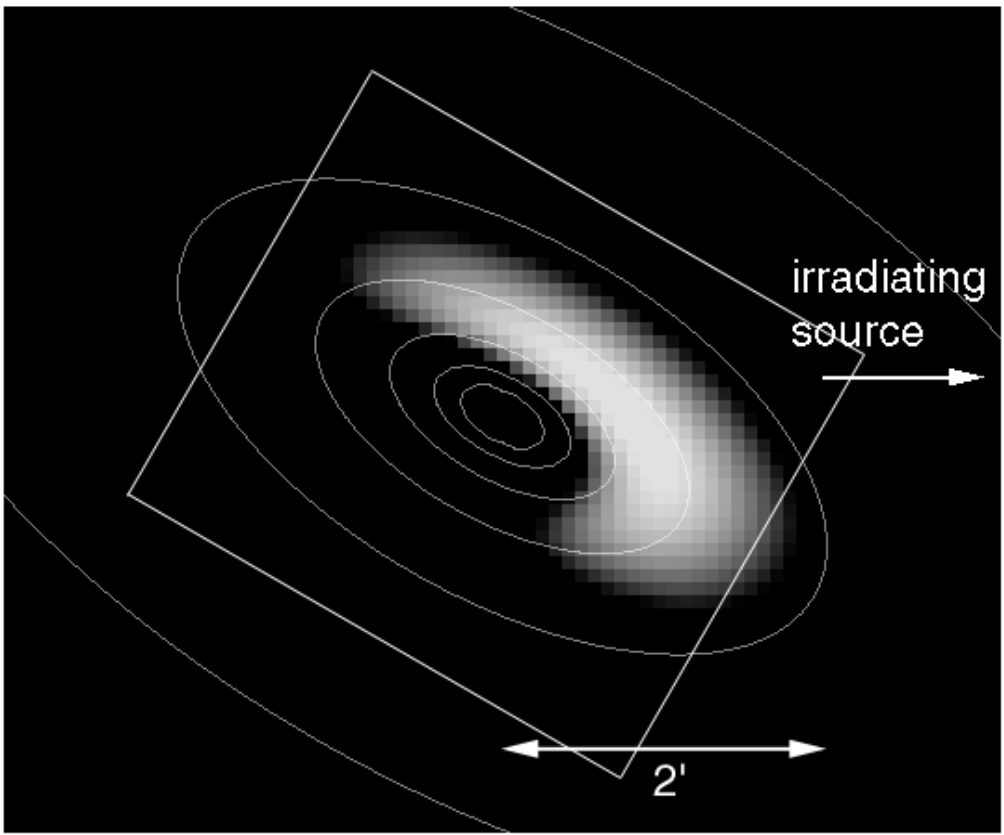} 
\caption[f3.ps]
{The simulated brightness distribution of the 6.4-keV
line overlaid on the contours of the molecular gas model. An
irradiating source is assumed to be in the Galactic center direction
(the right hand side; shown by the arrow). The solid
rectangle is the same region as that of Figure~1. }
\end{figure}

\clearpage

\begin{deluxetable}{ccccc}
\tabletypesize{\small}
\tablecaption{Best-Fit Results of the Combined Spectrum of Point Sources}
\tablehead{
\colhead{}   		&\colhead{$N_{\rm H}$\tablenotemark{a}} 
&\colhead{$kT$\tablenotemark{b}} &\colhead{$F_{\rm X}$\tablenotemark{c}}
&\colhead{$L_{\rm X}$\tablenotemark{d}}\nl
\colhead{} &\colhead{($10^{23}$ H cm$^{-2}$)}
&\colhead{(keV)} &\colhead{($10^{-5}$ph s$^{-1}$ cm$^{-2}$)}
&\colhead{($10^{33}$ erg s$^{-1}$)}
}

\startdata
&1.9$^{+0.6}_{-0.4}$ &8.5$^{+9.3}_{-3.7}$ 
&1.5$^{+0.5}_{-0.2}$ & $2.9^{+1.1}_{-0.8}$ \nl
\tableline
Reduced $\chi^2$ (d.o.f.)	&&&&1.02 (12) \nl
\enddata

\tablecomments{Errors are at 90\% confidence level.}
\tablenotetext{a}{Hydrogen column density.}
\tablenotetext{b}{Temperature of a thin thermal plasma.}
\tablenotetext{c}{Flux (no correction of absorption) in the 2--10 keV
 band.}
\tablenotetext{d}{Absorption corrected luminosity in the 2--10 keV band} 
\end{deluxetable}

\clearpage

\begin{deluxetable}{cccccc}
\tabletypesize{\small}
\tablecaption{Best-Fit Results of the Diffuse X-Ray Spectrum}
\tablehead{
\colhead{} & \multicolumn{4}{c}{Continuum}&\colhead{}\\
\cline{2-5}\nl
\colhead{}&
\colhead{$N_{\rm H}$\tablenotemark{a}}
&\colhead{$N_{\rm Fe}$\tablenotemark{b}} &\colhead{$\Gamma$\tablenotemark{c}}
&\colhead{$F_{\rm X}$\tablenotemark{d}}
&\colhead{}
\nl
\colhead{}&\colhead{($10^{23}$ H cm$^{-2}$)}&\colhead{($10^{19}$ Fe cm$^{-2}$)}
&\colhead{} &\colhead{($10^{-4}$ph s$^{-1}$ cm$^{-2}$)}
&\colhead{}
}
\startdata
&8.8 $^{+2.0}_{-1.5}$ &3.4 $^{+3.6}_{-2.2}$ &2.0 (fixed) 
&1.2 $^{+0.2}_{-0.1}$ \nl
\tableline
\tableline
& \multicolumn{4}{c}{Iron Lines}& Total\nl
& \multicolumn{2}{c}{K$\alpha$} &\multicolumn{2}{c}{K$\beta$} 
&\nl
\cline{2-5}\nl
{}&{$E$\tablenotemark{e}}
&{$F_{\rm X}$\tablenotemark{f}}&
{$E$\tablenotemark{g}}&{$F_{\rm X}$\tablenotemark{h}}
&$L_{\rm X}$\tablenotemark{i}
\nl
{}&{(keV)} &{($10^{-5}$ph s$^{-1}$ cm$^{-2}$)}
&{(keV)} &{($10^{-5}$ph s$^{-1}$ cm$^{-2}$)}
&($10^{35}$ erg s$^{-1}$)
\nl
\tableline\nl
&6.38$^{+0.02}_{-0.01}$ &5.6 $^{+0.5}_{-0.5}$ 
&7.04$^{+0.02}_{-0.01}$ &1.1 $^{+0.1}_{-0.1}$ 
&$1.0^{+0.4}_{-0.2} $
\nl
\tableline
Reduced $\chi^2$ (d.o.f.)		&&&&&1.08 (19) \nl
\enddata
\tablecomments{Errors are at 90\% confidence level.}
\tablenotetext{a}{Hydrogen column density.}
\tablenotetext{b}{Iron column density.}
\tablenotetext{c}{Photon index of a power-law model for the continuum spectrum.}
\tablenotetext{d}{Flux (no correction of absorption) in the 4--10 keV
 band.}
\tablenotetext{e}{Center energy of iron K$\alpha$ line.}
\tablenotetext{f}{Absorption corrected flux of iron K$\alpha$ line.}
\tablenotetext{g}{Center energy of iron K$\beta$ line. This value is
 assumed to be 1.103 $\times E_{\rm K\alpha}$ (see text).}
\tablenotetext{h}{Absorption corrected flux of iron K$\beta$ line. This
 value is assumed to be 0.1348 $\times F_{\rm K\alpha}$ (see text).}
\tablenotetext{i}{Absorption corrected luminosity in the 4--10 keV band.}
\end{deluxetable}

\end{document}